# DISCUSSION OF: STATISTICAL ANALYSIS OF AN ARCHEOLOGICAL FIND—SKEPTICAL COUNTING CHALLENGES TO AN ARCHAEOLOGICAL FIND

BY SHEILA M. BIRD

*MRC Biostatistics Unit*

The New Testament (NT) tomb in East Talpiyot, Jerusalem was discovered around Easter in 1980. Its surveyors at the time included Amos Kloner, whose 1980 PhD thesis was entitled "Tombs and Burials in the Second Temple Period," a topic on which he continued to publish for at least the next 15–20 years. Why did such a scholar not seize avidly the apparent historical opportunity that fell to his lot?

The tomb's excavator, Yosef Gath of the Department of Antiquities and Museums, died (date not specified) of heart failure not long after completing his work at the site. Upon completion of salvage excavations, "such bone material as remained was reburied" in accordance with Jewish ritual law. How much bone material remained? I assume that the orthodox rabbinate properly records reburials? Coincidentally, the NT tomb was discovered just as Sir Alec Jeffreys (1978–84, in Leicester, UK) was discovering DNA fingerprinting [see http://genome.wellcome.ac.uk/doc_wtd020877.html and Jeffreys, Wilson and Thein (1985)]. Some DNA analysis has been essayed, which Feuerverger side-steps. Shimon Gibson's archaeological drawings at the time of excavation indicated 10 ossuaries.

Ossuaries from the NT tomb were taken into the State of Israel Collections, but not until 1996 was it realized that records of the Israel Antiquities Authority (IAA) show only nine as having been received by it. Counting them all out and counting them all in, as famously reported by a UK journalist in the Falklands War, was inexplicably lax.

According to a 1994-published interpretation by authority Rahmani, and endorsed in 1996 by Kloner, six were found to have such Hebrew inscriptions as "Marya," "Yoseh," "Yeshua son of Yehosef," "Yehuda son of Yeshua," "Matya"...or Greek inscription of "Marmamene [diminutive] who is also called Mara." Attributions of authority are notoriously fickle: Rahmani had

---

also interpreted Mary and Joseph as the parents of Yeshua and grandparents of Yehuda. Feuerverger argues that, if Rahmani is correct in this interpretation, then the tombsite cannot be that of the NT family. The heretical alternative (which ancient religious authorities may have disavowed, or been unaware of) of Yeshua's having had a son by Mara is not admitted as a scientific (prior) consideration.

Rahmani's interpretation of the ossuaries' inscriptions is clearly a valid reason for the NT tomb's having not roused in the 1980s such titanic excitement as has since been engendered (http://www.theherald.co.uk/features/features/display.var.1226604.0.0.php).

As a practical statistician, my first set of sceptical questions therefore relates to the exact chronology of the tomb's discovery and excavation, the reburial of bone material (and its subsequent retrieval for DNA analysis), the registration(s) of ossuaries and deciphering of inscriptions, and the time-trail of interpretations of those inscriptions versus the publication of said interpretations.

Let me illustrate chronology by a controversy in the UK press in early January 2008 (see http://media.newscientist.com/data/pdf/press/2637/263711.pdf and http://www.guardian.co.uk/science/2008/jan/03/medicalresearch.agriculture) which surrounds the publication in December 2007 of a case-study that was submitted to Archives in Neurology [Mead et al. (2007)], an American journal, in February 2006. It concerns a 39-year old woman who died in 2000, 14 months after clinical onset of disease that was ascribed to sporadic CJD (despite atypical findings at post-mortem). Of particular note were: (a) that she was valine homozygous at codon 129 of the prion protein, and (b) that molecular analysis of cerebellar tissue demonstrated a novel $PrP^{Sc}$ type similar to that seen in vCJD. The authors reported that transmission studies were underway. This lady, were she the first clinical case of vCJD in a patient who is *not* methionine homozygous at codon 129 of the prion protein, would be as important as a first as was human-to-human, blood-borne transmission of vCJD, which merited parliamentary announcement in UK. Mysterious, therefore, were the up-to-seven-year delay in publication, failure to cite when transmission studies in mice had begun, and the authors' apparent caution that this was, in fact, *not* vCJD. Only a limited post-mortem had been permitted so that lymphoid tissue, such as from spleen and appendix, were not available for testing. The patient had a tonsillectomy but at a date and hospital unspecified; and some of the molecular techniques used were relatively recent. Transmission studies had been underway for some time so that preliminary results from them may indeed have underpinned the authors' caution. I recount this cautionary tale for two reasons: first, to illustrate that statisticians may need a hinterland of subject-matter knowledge to identify the critical questions to ask before proceeding to inference . . . and, secondly, because it would be epidemiologically shocking if, for seven years, UK had overlooked vCJD in a clinical case who was valine–valine and, accordingly, the time-trail might point to pathological or molecular lacunae that needed to be plugged in UK's, European and world-wide CJD surveillance.

Let me end with the other conundrum: the missing or stolen ossuary from the NT tomb—an archaeological, if not criminal, travesty. Was an ossuary inscribed "James son of Joseph brother of Jesus" and in the possession of a private Israeli antiquities collector under prosecution for alleged forgery of *part* of said inscription from the NT tomb? Feuerverger notes that, due to the Sabbath, the NT tomb was left open from Friday afternoon to Sunday morning in the four-day period of 28–31 March 1980. He speculates that investigating archaeologists were unlikely to have missed a seventh inscription (even prior to their having been "cleaned up") on the 10 ossuaries they'd located. Thus, if the "James" ossuary indeed came from the NT tomb, it would have to have been an 11th that the investigating archaeologists had somehow overlooked. That conveniently leaves the "missing" 10th ossuary as uninscribed. This line of argument is flimsy, but so too is it extraordinary to me that such antiquities were: (a) left open, (b) inaccurately curated, and (c) long under-rated as potentially newsworthy... unless scholars had indeed posed critical questions, and deployed DNA or other scientific techniques, that have unveiled more context than the problem posited, somewhat mysteriously, to investigator Feuerverger to cast statistical light on. Know thine enemy (bias).

MRC Biostatistics Unit
Institute of Public Health
University Forvie Site
Robinson Way
Cambridge CB2 2SR
United Kingdom
E-mail: sheila.bird@mrc-bsu.cam.ac.uk